\documentclass[aps,prd,onecolumn,nofootinbib,superscriptaddress]{revtex4}
\pdfoutput=1
\usepackage{float}
\usepackage{graphicx}%Include figure files
\usepackage{amsmath}
\usepackage{amsfonts}
\usepackage{amssymb,ulem}
\usepackage{mathrsfs}
\usepackage{color}%
\usepackage{dcolumn}
\usepackage{subfigure}
\usepackage{pdfpages}
\usepackage{multirow}
\usepackage{MnSymbol,wasysym}
\usepackage{braket,diagbox}
\usepackage{eurosym}
\usepackage{euscript}
\usepackage[usenames,dvipsnames,svgnames]{xcolor}

\newcommand{\RNum}[1]{\uppercase\expandafter{\romannumeral #1\relax}}
\usepackage[colorlinks=true,linkcolor=blue,urlcolor=blue,filecolor=black,citecolor=red,
pdfstartview=FitV,pdftitle={},pdfsubject={},pdfkeywords={},pdfpagemode=None,bookmarksopen=true]{hyperref}

\begin{document}
\baselineskip=0.5 cm

\title{Probing a Lorentz-violating parameter from orbital precession of the S2 star around the galactic centre supermassive black hole}

\author{Qi Qi}
\email{qiqiphy@163.com}
\affiliation{Center for Gravitation and Cosmology, College of Physical Science and Technology, Yangzhou University, Yangzhou, 225009, China}

\author{Yu Sang}
\email{sangyu@yzu.edu.cn}
\affiliation{Center for Gravitation and Cosmology, College of Physical Science and Technology, Yangzhou University, Yangzhou, 225009, China}

\author{Xiao-Mei Kuang}
\email{xmeikuang@yzu.edu.cn (corresponding author)}
\affiliation{Center for Gravitation and Cosmology, College of Physical Science and Technology, Yangzhou University, Yangzhou, 225009, China}

\date{\today}

\begin{abstract}
\baselineskip=0.5 cm
Testing Lorentz symmetry in strong gravitational fields provides a promising probe of extensions to general relativity. The supermassive black hole Sgr~A* and the orbit of the S-stars offer a laboratory for such tests in a regime beyond weak field limit. We analyze the S2 orbital data focusing on the Schwarzschild-like black hole within bumblebee gravity, where deviations from general relativity are encoded in a single Lorentz-violating parameter $\ell$. Using a full 14-dimensional Markov Chain Monte Carlo analysis under uniform and Gaussian priors, we obtain $\ell = {-8.01 \times 10^{-5}}^{+2.77 \times 10^{-4}}_{-2.09 \times 10^{-4}} $ and $\ell =  {1.00 \times 10^{-5}}^{+2.90 \times 10^{-4}}_{-2.91 \times 10^{-4}} $ at $1\sigma$ confidence level, respectively. These constraints are about three orders of magnitude tighter than those from Event Horizon Telescope imaging of Sgr~A*. 
\end{abstract}

\maketitle
\tableofcontents
\newpage

\section{Introduction}
Supermassive black holes (SMBHs), with masses in the range $10^5$-$10^9M_{\odot}$, are widely believed to reside at the centers of galaxies \cite{Gultekin:2009qn,Genzel:2010zy}. The Event Horizon Telescope (EHT) has provided compelling electromagnetic evidence for the existence of SMBHs in M87 and at the Galactic center, Sagittarius A* (Sgr~A*) \cite{EventHorizonTelescope:2019dse,EventHorizonTelescope:2022wkp}. These horizon-scale images offer direct insights into the spacetime geometry and accretion environments of black holes, enabling stringent tests of general relativity (GR) in the strong field regime \cite{Bambi:2019tjh,EventHorizonTelescope:2019pgp,EventHorizonTelescope:2019ggy,EventHorizonTelescope:2020qrl,EventHorizonTelescope:2021srq,EventHorizonTelescope:2021dqv,EventHorizonTelescope:2022xqj}. In parallel, gravitational wave observations are expected to provide an independent and complementary probe of SMBHs in the near future, particularly through the planned space-based interferometers such as the Laser Interferometer Space Antenna (LISA) \cite{LISA:2017pwj}, Taiji \cite{Hu:2017mde}, and TianQin \cite{TianQin:2015yph}.

The dynamics of stars orbiting around a SMBH provide a powerful avenue for probing new physics, particularly in light of the rapidly improving observational capabilities. Two of the most prominent observational programs are conducted by the Keck Observatory \cite{Ghez:2003qj} and the Very Large Telescope (VLT) with the GRAVITY instrument \cite{Gillessen:2017jxc,GRAVITY:2021xju}, which have monitored the motions of several stars around Sgr~A* for more than two decades. In particular, near-infrared observations from the Keck and GRAVITY/VLTI collaborations have reconstructed the orbits of the so-called S-stars, most notably the S2 star, with remarkable precision \cite{Gillessen:2017jxc,GRAVITY:2018ofz,GRAVITY:2020gka}. The S2 star follows a highly eccentric ($e\simeq0.88$) orbit with a period of about 16 years, reaching a pericentre distance of $\sim120 \mathrm{AU}$. Current observations have confirmed the gravitational redshift and the pericentre precession predicted by GR within observational uncertainties \cite{Do:2019txf,GRAVITY:2021xju}. Owing to the extreme gravitational environment and the high precision of the astrometric and spectroscopic data, the S2 orbit serves not only as a potential probe of fundamental physics, such as the distribution and nature of dark matter \cite{Buckley:2017ijx,Nampalliwar:2021tyz,Tsai:2021irw,Yuan:2022nmu}, but also as a natural laboratory for testing GR and exploring possible deviations from it in alternative theories of gravity \cite{Shaymatov:2023jfa,DellaMonica:2021xcf,DellaMonica:2022eeg,Li:2022rjv,Yan:2022fkr,Zhang:2024fpm,deMartino:2021daj,Fernandez:2023kro,Li:2025qcv,GRAVITY:2023azi,GRAVITY:2023cjt,DellaMonica:2023dcw}.

Lorentz symmetry is a cornerstone of modern physics, underlying both the Standard Model of particle physics and GR. It ensures the invariance of physical laws under local Lorentz transformations and plays a fundamental role in the equivalence principle. Nevertheless, several candidate theories of quantum gravity, such as string theory, loop quantum gravity and noncommutative geometry, suggest that Lorentz invariance may not be exact, but could instead be spontaneously or dynamically broken at the Planck scale \cite{Kostelecky:1988zi, Amelino-Camelia:2008aez}. This possibility has motivated extensive theoretical and experimental efforts to search for possible signatures of Lorentz symmetry breaking (LSB) across a wide range of energy and curvature regimes \cite{Mattingly:2005re, Liberati:2013xla}.

A framework for quantifying possible violations of Lorentz invariance is provided by the standard model extension (SME), developed by Kosteleck \textit{et al.} \cite{Colladay:1998fq, Kostelecky:2003fs}. Within the SME, one of the simplest realizations in the gravitational sector is the bumblebee gravity model, in which a vector field acquires a nonzero vacuum expectation value, thereby inducing spontaneous LSB. The action for a single bumblebee field
$B^{\mu}$ coupled to gravity can be written as \cite{Kostelecky:1988zi,Kostelecky:1989jw,Bluhm:2004ep}
\begin{equation} \label{eq:Action}
S_{B}=\int{d^4x} \mathscr{L}_{B} =\int{d^4x} \sqrt{-g}[\frac{1}{16\pi G} R+\frac{\xi}{16\pi G}B^{\mu}B_{\nu}R_{\mu\nu}-\frac{1}{4} B_{\mu\nu}B^{\mu\nu}-V(B^{\mu})],
\end{equation}
where $\xi$ is a coupling constant characterizing the nonminimal  interaction between the bumblebee field and gravity, $B_{\mu\nu} = \partial_{\mu}B_{\nu}-\partial_{\nu}B_{\mu}$ is the field strength, and $V$ is the potential responsible for triggering spontaneous LSB by enforcing a nonzero vacuum expectation value of $B^{\mu}$.  Casana \textit{et al.} \cite{Casana:2017jkc} obtained the first exact Schwarzschild-like black hole solution in this framework  by considering a potential with the form
\begin{equation} \label{eq:Vpo}
V \equiv V(B^{\mu}B_{\mu}\pm b^{2}),
\end{equation}
which enforces the condition $B^{\mu}B_{\mu}=\mp b^{2}$. As a result,  the bumblebee field acquires a background value $\langle B^\mu \rangle = b^\mu$ with constant norm $b^{\mu}b_{\mu}=\mp b^{2}$, leading to spontaneous breaking of local Lorentz invariance.
Fixing  the bumblebee field  at its vacuum expectation value as  $B_{\mu}=b_{\mu}$ and considering  a radial spacelike bumblebee background $b_{\mu}=(0,b_{r}(r),0,0)$, the resulting static and spherically symmetric Schwarzschild-like metric takes the form
\begin{eqnarray}
ds^{2} &=&-g_{tt}dt^2+g_{rr}dr^2+r^2(d\theta^2+\sin^2\theta d\phi^2)\label{eq:general metric}\\
 &=&-\Big(1-\frac{2M}{r}\Big)dt^{2} + (1+\ell)\Big(1-\frac{2M}{r}\Big)^{-1}dr^{2} + r^{2}(d\theta^{2}+\sin^{2}\theta\,d\phi^{2}).\label{eq:bumblebee_metric_sec2}
\end{eqnarray}
The dimensionless parameter $\ell \equiv \xi b^2$ originates from the nonzero vacuum expectation value of the bumblebee field and the nonminimal coupling to gravity. Physically, $\ell$ measures the strength of spontaneous LSB in this model and should satisfy $\ell>-1$ to ensure the Lorentz signature of the metric. In what follows, we shall assume $|\ell|\ll1$ and work perturbatively in $\ell$.
Inspired by this solution, a variety of Lorentz-violating black hole spacetime and their observable consequences have been extensively explored,  including  slowly rotating Kerr-like black holes \cite{Ding:2019mal} and their accretion images \cite{Ovgun:2018ran,Kuang:2022xjp,Xu:2025iwg},  Kerr-Sen-like solution and its shadow \cite{Jha:2020pvk}, as well as implications for gravitational wave signals \cite{Xu:2022frb,Liang:2022gdk,Lai:2025nyo}. Further generalizations and phenomenological studies can be found in \cite{Delhom:2019wcm,Filho:2022yrk,Liu:2022dcn,Liu:2024axg,Bailey:2025oun,Chen:2025ypx,Li:2025bzo,Li:2025tcd,Shi:2025hfe} and references therein.

Testing Lorentz symmetry in gravitational fields near compact objects provides an essential complement to laboratory and cosmological experiments. The SMBH at the Galactic Centre, Sgr~A*, offers a  natural laboratory for such investigations. In this work, we explore potential signatures of LSB in the orbital dynamics of the S2 star. Specifically, we consider the Schwarzschild-like black hole solution \eqref{eq:bumblebee_metric_sec2} in bumblebee gravity, which encodes spontaneous LSB through a dimensionless parameter $\ell$. This parameter modifies the effective gravitational potential and, consequently, the relativistic orbital precession of test particles.

To this end, we derive the geodesic equations and obtain analytic expressions for the pericentre precession angle, expanded to leading order in the Lorentz-violating parameter $\ell$. These theoretical predictions are then confronted with the latest astrometric and spectroscopic observations of the S2 star. By performing a Markov Chain Monte Carlo (MCMC) analysis with two sets of priors, namely uniform and Gaussian priors, over a 14-parameter space, we place constraints on the allowed range of $\ell$ and assess the consistency of the inferred precession rate with the prediction of GR.

%The constraints on LSB bound in the strong-gravity regime obtained from the S2 star data provide independent results from SMBH rather than  Solar System and pulsar timing experiments. Furthermore, these findings complement existing constraints obtained from gravitational-wave dispersion analyses \cite{LIGOScientific:2019fpa,Shao:2014oha}, thereby extending the empirical foundation of local Lorentz invariance to the regime of supermassive black holes.

The paper is organized as follows.
In section \ref{sec:metrics_geodesics}, we present a very brief review on the geodesic motion of a massive test particle and calculate its pericentre precession in the Schwarzschild-like black hole spacetime. In section
\ref{sec: precession orbit}, we first briefly describe the publicly available astrometric and spectroscopic data used in this paper, and then we build the orbital model and contrast it with the data
used in an MCMC simulation. We then perform the MCMC analysis with both uniform prior and Gaussian prior, and discuss the main results of our analysis in section \ref{sec:MCMC}. Section \ref{sec:conclusion} contributes to our closing remarks.

\section{Geodesic motion and leading-order pericentre precession}
\label{sec:metrics_geodesics}

In this section, we partly review the process in \cite{Casana:2017jkc} to derive the geodesic equations for timelike test particles and compute the leading-order correction to the pericentre precession induced by the Lorentz-violating parameter $\ell$. We set $G=c=1$. Since the above metric \eqref{eq:general metric} or \eqref{eq:bumblebee_metric_sec2} has spherical symmetry, we can safely consider the
motions of the massive particles in the equatorial plane with $\theta = \pi/2$.
The Lagrangian $\mathscr{L}=\tfrac12 g_{\mu\nu}\dot x^\mu\dot x^\nu$ gives two Killing constants, the energy and angular momentum per unit rest mass, as
\begin{align}
E &\equiv -g_{tt}\dot t, \label{eq:E_def_sec2}\\
L &\equiv g_{\phi\phi}\dot\phi = r^{2}\dot\phi, \label{eq:L_def_sec2}
\end{align}
where the dot denotes the derivative with respect to the proper time $\tau$. Then for a timelike particle, the normalization $g_{\mu\nu}\dot x^\mu\dot x^\nu = -1$ yields the radial geodesic equation,
\begin{equation}
\label{eq:radial_tau_sec2}
g_{rr}\dot r^{2} = E^{2} - V_{\rm eff}(r),~~~~\text{with}~~~~
V_{\rm eff}(r) \equiv -g_{tt}\Big(1+\frac{L^{2}}{r^{2}}\Big).
\end{equation}
Combining \eqref{eq:L_def_sec2} and \eqref{eq:radial_tau_sec2}, we have
\begin{equation}
\label{eq:radial_phi_sec2}
\left(\frac{dr}{d\phi}\right)^{2} = \frac{r^{4}}{L^{2}}\frac{1}{g_{rr}}\Big[(E^{2}-V_{\rm eff}(r)\Big].
\end{equation}
Introducing the new coordinate $u \equiv 1/r$ and taking the derivative with respect to $\phi$ on \eqref{eq:radial_phi_sec2} further gives a second-order equation as
\begin{equation}
\label{eq:second_order}
\frac{d^2u}{d\phi^2}=\frac{1}{2L^2}\Big[-\frac{1}{g^2_{rr}} \frac{dg_{rr}}{du}(E^2 -V_{\rm eff})+\frac{1}{g_{rr}} (-\frac{dV_{\rm eff}}{du}) \Big].
\end{equation}

\begin{figure} [h]
{\centering
\includegraphics[width=3 in]{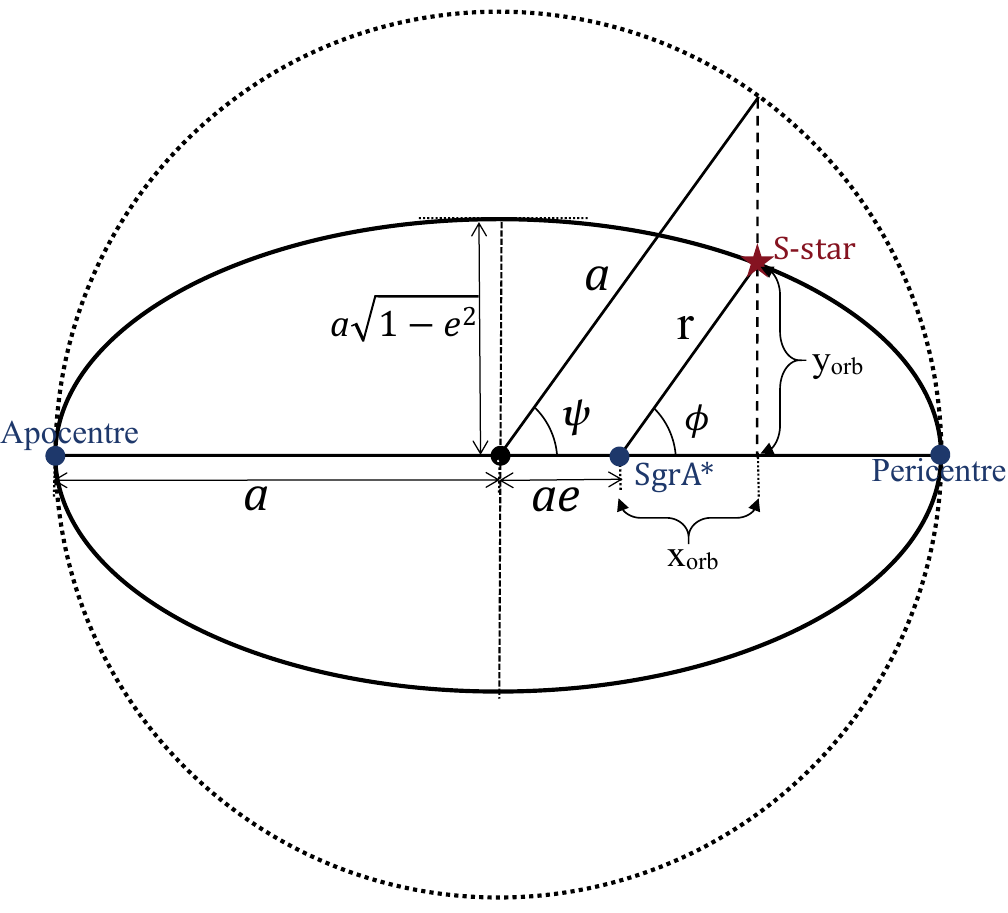} \hspace{0.5cm}
   \caption{ This figure shows the orbit of the S-star and its orbital elements. The solid ellipse represents the trajectory of the S-star, where $a$ is the semi-major axis, $e$ is the eccentricity, $\psi$ is the eccentric anomaly, $\phi$ is the true anomaly, and $r$ is the distance between the S-star and the SMBH Sgr~A*.}   \label{fig:CCT}}
\end{figure}

Recalling the metric \eqref{eq:bumblebee_metric_sec2}, the formulas \eqref{eq:E_def_sec2} and \eqref{eq:L_def_sec2} can be rewritten as
\begin{equation}
E=(1-2Mu) \dot{t},~~L=1/u^2 \dot{\phi},
\end{equation}
and the first order radial equation \eqref{eq:radial_phi_sec2} reduces to
\begin{equation}
\label{eq:bum_integ}
\left(\frac{du}{d\phi}\right)^{2} =\frac{1}{1+\ell} \Big[ \frac{E^2-1}{L^2}+\frac{2M}{L^2}u+2Mu-u^2 \Big].
\end{equation}
The above scenario describes the periastron precession of the star's motion orbiting Sgr~A*, which is depicted in Fig. \ref{fig:CCT}. It is obvious that the particle's trajectory can be traced by
\begin{equation}\label{eq:x(Psi)}
(\mathsf{x_{\mathrm{orb}}},\mathsf{y_{\mathrm{orb}}}) = (a (\cos \psi-e), a \sqrt{1-e^2} \sin \psi),
\end{equation}
where $e$ is the orbital eccentricity and $a$ is the semi-major axis. $\psi$ is the eccentric anomaly, and the pericentre and apocentre are given by  $\psi=\phi=0$ and $\psi=\phi=\pi$, respectively. At these two points, we have
\begin{eqnarray}\label{eq:r(Psi)}
r_{a}=\sqrt{\mathsf{x^2_{\mathrm{orb,a}}}+\mathsf{y^2_{\mathrm{orb,a}}}} = a(1+e),\\
r_{p}=\sqrt{\mathsf{x^2_{\mathrm{orb,p}}}+\mathsf{y^2_{\mathrm{orb,p}}}} = a(1-e),
\end{eqnarray}
where $r_{a}$, $\mathsf{x_{\mathrm{orb,a}}}$ and $\mathsf{y_{\mathrm{orb,a}}}$ are the position coordinates of the S-star (massive particle) at the apocentre, while $r_{p}$, $\mathsf{x_{\mathrm{orb,p}}}$ and $\mathsf{y_{\mathrm{orb,p}}}$ are those at the pericentre.

%then $u_a=1/r_a=1/1+e$,$u_p=1/r_p=1/1-e$ and the radial geodesic equation satisfies $du/d\phi =0$, then eq\eqref{eq:bum_integ} simplifies to:
Moreover, considering that $du/d\phi =0$ at both pericentre and apocentre, \eqref{eq:bum_integ} at these points suggests
\begin{equation}\label{eq:bum_integ-}
\frac{E^2-1}{L^2} +\frac{2M}{L^2} u_{p,a} + 2Mu^3_{p,a} - u^2_{p,a} =0,
\end{equation}
where $u_a=1/r_a=1/(a(1+e))$, $u_p=1/r_p=1/(a(1-e))$. Further subtracting the two equations in \eqref{eq:bum_integ-}, we can cancel the energy $E$ and obtain the expression of angular momentum $L$. For Keplerian motion, the orbital scale is determined by the semi-major axis $a$ while the relativistic length scale is given by the mass parameter $M$. Since the orbital size $a$  is much larger than the gravitational radius $M$, we have $M/a \ll 1$ \cite{Will:2014kxa, poisson2014gravity}. Then, retaining only the leading Keplerian contribution, the angular momentum reduces to
\begin{equation} \label{eq:lequ}
L^2=Ma(1-e^2)+ \mathcal{O}(\frac{M^2}{a^2}) \simeq Ma(1-e^2).
\end{equation}

Subsequently, the second order equation \eqref{eq:second_order} can be simplified into
\begin{equation}
\label{eq:bum_integdouble}
\frac{d^2u}{d\phi^2} =\frac{1}{1+\ell} \Big[\frac{M}{a(1-e^2)}+3Mu^2-u \Big] .
\end{equation}
In order to solve \eqref{eq:bum_integdouble} perturbatively,  we introduce the small post-Newtonian (PN) parameter $\epsilon \equiv3M^2/L^2 \simeq3M/(a(1-e^2)) \ll1$ and set $|\ell|\ll1$. Solving the equation order by order with the initial conditions $u(0)=u_p$ and $\dot{u}(0)=0$, we can obtain the solution in the form $u\simeq u^{(0)}+\epsilon u^{(1)}$, with
\begin{align}
u^{(0)} &=\frac{1}{a(1-e^2)} \Big[1+e \cos(\frac{\phi}{\sqrt{1+\ell}}) \Big], \label{bum_solution}\\
u^{(1)} &\simeq\frac{1}{a(1-e^2)} e \frac{\phi}{\sqrt{1+\ell}} \sin\Big(\frac{\phi}{\sqrt{1+\ell}}\Big) + \frac{1}{a(1-e^2)}\Big[(1+\frac{e^2}{2})-\frac{e^2}{6}\cos\Big(\frac{2 \phi}{\sqrt{1+\ell}}\Big) \Big], \label{eq:bum_soludouble}
\end{align}
where $u^{(0)}$ is the zeroth-order solution while $u^{(1)}$ is the first-order PN correction. And, the combined solution can be further approximated into the form
\begin{equation}
u\simeq u^{(0)}+\epsilon u^{(1)} = \frac{1}{a(1-e^2)} \Big[1+e \cos\Big(\frac{\phi (1-\epsilon)}{\sqrt{1+\ell}}\Big) \Big].
\end{equation}
Thus, when $\phi (1+\epsilon)/\sqrt{1+\ell}=2 \pi n$, $u$ is at the periastron and the orbital period $\Phi$ is
\begin{equation}
\label{eq:bum_or_period}
\Phi = \frac{2\pi \sqrt{1+\ell}}{1+\epsilon}=2 \pi +\Delta\phi.
\end{equation}
Here the orbital precession $\Delta\phi$ for the Schwarzschild-like black hole \eqref{eq:bumblebee_metric_sec2} can be solved by taking the lower-order terms in the expansion of the small parameters $\ell$ and $\epsilon$ as
\begin{equation}
\label{eq:bum_orbit_period}
\Delta\phi\simeq 2\pi (1+\epsilon+\frac{\ell}{2})-2\pi=2\pi\epsilon+\pi \ell.
\end{equation}
It is noted that when $\ell=0$, the above formula reduces to the standard GR precession,
\begin{equation} \label{eq:GRprece}
\Delta\phi_{\rm GR}=2\pi\epsilon=\frac{6 \pi M}{a(1-e^2)}.
\end{equation}

\section{Data and data analysis of the S2 star}\label{sec: precession orbit}
The S2 star is a  nuclear stellar cluster orbiting the radio source Sgr~A* at the Galactic Center. Its orbit is well measured, with an orbital period of about $16$ years, a semi-major axis of approximately $970 \mathrm{AU}$, and a high eccentricity of $e \simeq 0.88$ \cite{Gillessen:2017jxc,Do:2019txf,GRAVITY:2018ofz}. Over the past three decades, the GRAVITY Collaboration has continuously monitored S2 star, obtaining highly precise astrometric and spectroscopic data. Recently, they have also measured key relativistic effects along S2 star's orbit, including the gravitational redshift and the pericentre advance of $\delta\phi \simeq 12'$ per orbital revolution. All of these observations remain fully consistent with the predictions of GR.

At the same time, the precision of the S2 dataset provides a powerful avenue for testing small deviations from GR in the strong-gravity environment of the SMBH. In this study, we use the publicly available S2 star observations to constrain the spacetime metric involving Lorentz-violating extension. The resulting bounds can then be directly translated into limits on the corresponding Lorentz-violating parameter in the current theoretical framework.

\subsection{Dataset}
We use the publicly available astrometric and spectroscopic data for the S2 star, which have been collected over the past decades. The full dataset is divided into three components: astrometric positions, radial velocities, and  the orbital precession. A brief summary of these data is provided below.

\begin{figure} [h]
{\centering
\includegraphics[width=3.2in]{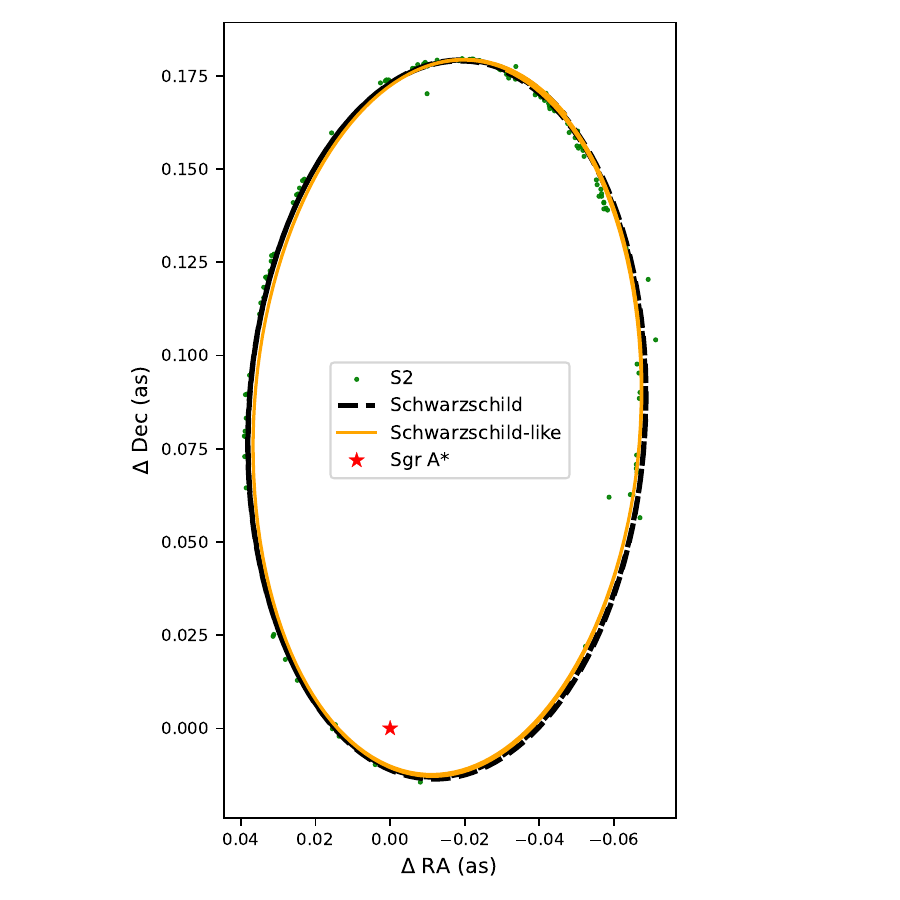}
\includegraphics[width=3.8in]{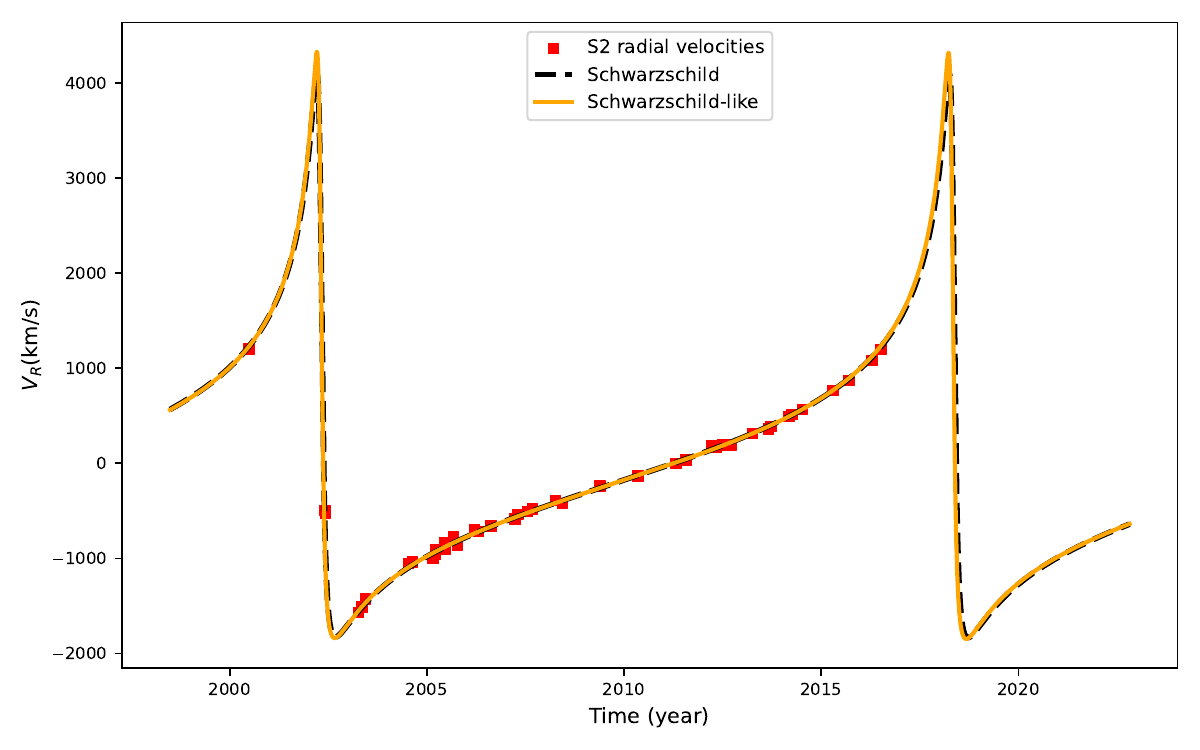}
   \caption{{\bf Left:} The theoretically calculated orbit and the observational data points of astrometric positions of the S2 star. The black dashed curve shows the orbit predicted in the Schwarzschild black hole. The orange solid curve corresponds to the prediction obtained in the Schwarzschild-like black hole. The green circulars represent the observed astrometric positions. Sgr~A* is located at the coordinate origin. {\bf Right:} The variation of the radial velocity of the S2 star over time. The black dashed curve represents the prediction of the Schwarzschild black hole while the solid orange curve shows the result obtained from the Schwarzschild-like case. The red squares represent the observed radial velocities. }   \label{fig:OBC}}
\end{figure}

{\bf Astrometric positions (AP):} We employ 145 astrometric measurements of the sky-projected position of S2 spanning the period from 1992.224 to 2016.53 \cite{Gillessen:2017jxc}. These observations are divided into two groups according to the instruments used. Measurements prior to 2002 were obtained using the speckle-imaging camera SHARP at the ESO New Technology Telescope (NTT), with a typical astrometric accuracy of about $3.8~\mathrm{mas}$. From 2002 onward, observations were carried out with the adaptive-optics-assisted infrared camera NAOS+CONICA (NACO) at the Very Large Telescope (VLT), yielding a significantly improved accuracy of roughly $400~\mu\mathrm{as}$. These data correspond to the green dots in the left plot of Fig.~\ref{fig:OBC}.

{\bf Radial velocities:} We use 44 radial velocity measurements collected between 2000.487 and 2016.519 \cite{Gillessen:2017jxc}. Similar to the astrometric data, these measurements are separated into two groups. Data taken before 2003 were obtained using the adaptive-optics imager and spectrometer NIRC2 at the Keck Observatory, while those from 2003 onward were measured with the Spectrograph for INtegral Field Observations in the Near Infrared (SINFONI), an adaptive-optics-assisted integral field spectrograph at the VLT. These data are shown as red squares in the right plot of Fig.~\ref{fig:OBC}.

{\bf Orbital precession:} Besides the astrometric and radial velocity data, we also include the pericentre-advance measurement of S2 reported by the GRAVITY Collaboration \cite{GRAVITY:2020gka}. Although the full GRAVITY astrometric dataset is not publicly released, the orbital precession has been  measured as
\begin{equation}\label{eq:deltaPHI}
f_{\rm sp}=\frac{\Delta\phi}{\Delta\phi_{\rm GR}} = 1.10 \pm 0.19 .
\end{equation}
This result agrees with the prediction of GR within $1\sigma$, and rules out a Newtonian orbit at more than $5\sigma$. Including this relativistic precession could provide an additional high-precision constraint on alternative gravitational theories.

\subsection{Modeling the orbit with relativistic effects}
\begin{figure} [h]
{\centering
\includegraphics[width=3 in]{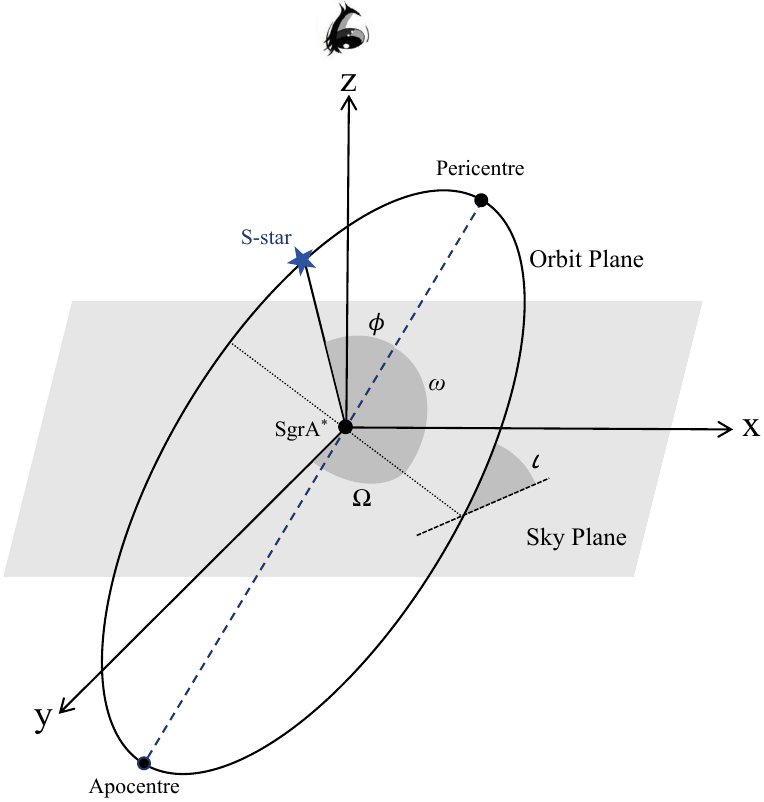} \hspace{0.5cm}
   \caption{The relationship between the orbital plane of the S-star and the observer's plane, along with the relevant Keplerian orbital elements. Here, the $Z$-axis coincides with the observer's line of sight, $\Omega$ is the longitude of ascending node, $\omega$ is the argument of pericentre (i.e., the angle from the ascending node to the pericentre measured in the orbital plane), the pentagram is the position of the S-star, and $\iota$ (the inclination) is the angle between the observer's plane and the orbital plane. }   \label{fig:CT}}
\end{figure}

By numerically integrating the geodesic equations with a given set of initial positions and velocities, we obtain the theoretical trajectory of the S2 star in the current black hole background. These computed positions naturally lie in the orbital plane of the star. In contrast, the astrometric observations discussed in the previous subsection record the apparent positions of S2 as projected onto the sky plane, i.e., the plane perpendicular to the line of sight of the observer. The geometric relation between these two planes is illustrated in Fig.~\ref{fig:CT}.

It is worth noting that, in both the orbital plane and the sky plane, the motion of S2 can be well approximated by a precessing elliptical orbit, even when small deviations from GR are included. To directly compare our theoretical predictions with the observed astrometric data, all relevant quantities must be expressed in the same observational frame. This requires projecting the numerically obtained orbit plane coordinates onto the sky plane through a sequence of rotations characterized by the orbital inclination $\iota$, the longitude of the ascending node $\Omega$, and the argument of pericentre $\omega$, see Fig.~\ref{fig:CT}.
This transformation is conveniently expressed using the Thiele-Innes constants, yielding
\begin{align}
X &= B \mathsf{x_{\mathrm{orb}}} + G \mathsf{y_{\mathrm{orb}}}, \label{eq:XX} \\
Y &= A \mathsf{x_{\mathrm{orb}}} + F \mathsf{y_{\mathrm{orb}}}, \label{eq:YY}\\
Z &= C \mathsf{x_{\mathrm{orb}}} + H \mathsf{y_{\mathrm{orb}}}, \label{eq:ZZ}
\end{align}
where $(\mathsf{x_{\mathrm{orb}}},\mathsf{y_{\mathrm{orb}}},\mathsf{z_{\mathrm{orb}}})$ are the Cartesian coordinates in the orbit plane, and $(X,Y,Z)$ represent the coordinates in the sky plane coordinate system, with $Z$ aligned along the observer's line of sight. The Thiele-Innes constants take the form
\begin{align}
A &= \cos\Omega\cos\omega - \sin\Omega\sin\omega\cos \iota, \
~~~B = \sin\Omega\cos\omega + \cos\Omega\sin\omega\cos \iota, \\
C &= -\sin\omega\sin \iota, \
~~~F = -\cos\Omega\sin\omega - \sin\Omega\cos\omega\cos \iota, \\
G &= -\sin\Omega\sin\omega + \cos\Omega\cos\omega\cos \iota, \
~~~H = -\cos\omega\sin \iota,
\end{align}
whose geometric meanings can be directly read off from Fig.~\ref{fig:CT}.

Similarly, the velocities are projected using the same transformation rules,
\begin{align}
V_X &= B \mathsf{v_{x,\mathrm{orb}}} + G \mathsf{v_{y,\mathrm{orb}}}, \\
V_Y &= A \mathsf{v_{x,\mathrm{orb}}} + F \mathsf{v_{y,\mathrm{orb}}}, \\
V_Z &= -(C \mathsf{v_{x,\mathrm{orb}}} + H \mathsf{v_{y,\mathrm{orb}}}),
\end{align}
where the negative sign in $V_Z$ ensures that the conventional definition of radial velocity remains positive when the star approaches the observer.

Moreover, to compare our theoretical predictions with the astrometric observations of S2, several relativistic and instrumental effects must be properly taken into account. These corrections ensure that the computed sky plane positions and radial velocities are expressed in the same observational reference frame as the measured data.

Firstly, since the observational reference frame is not perfectly aligned with the gravitational center, small offsets and linear drifts inevitably arise. Following \cite{Do:2019txf}, we model these effects by introducing four nuisance parameters, $x_0$, $y_0$, $v_{x0}$, and $v_{y0}$, such that the projected sky-plane positions are corrected as
\begin{eqnarray}
X &=& X(t_{\rm em}) + \mathsf{x_0} + \mathsf{v_{x0}} ,(t_{\rm em}-t_{\rm refer}),\\
Y &=& Y(t_{\rm em}) + \mathsf{y_0} + \mathsf{v_{y0}} ,(t_{\rm em}-t_{\rm refer}),
\end{eqnarray}
where $t_{\rm refer}$ denotes the reference epoch associated with the astrometric solution, and $t_{\rm em}$ is the emission time of the photons. The quantities $\mathsf{x_0}$ and $\mathsf{y_0}$ represent constant offsets, whereas the terms proportional to $\mathsf{v_{x0}}$ and $\mathsf{v_{y0}}$ encode linear drifts of the reference frame relative to the true dynamical center \cite{Gillessen:2017jxc}. In our analysis, we adopt $t_{\rm refer}=2009.2$ as used in \cite{plewa2015pinpointing}.

Secondly, the astrometric positions are measured at the observation time $t_{\rm obs}$, whereas the theoretical positions are naturally functions of the emission time $t_{\rm em}$. Because light propagates at a finite speed, the two times differ by the R\o mer delay. To leading order, this light-travel-time correction reads \cite{Do:2019txf,GRAVITY:2018ofz}
\begin{equation}
t_{\rm em} = t_{\rm obs} - \frac{Z(t_{\rm em})}{c},
\end{equation}
where $Z(t_{\rm em})$ is the line-of-sight coordinate obtained from  \eqref{eq:ZZ}, and $c$ is the speed of light. Solving this relation ensures that the orbit plane trajectory is consistently projected onto the sky plane at the appropriate observational epoch.

Thirdly, the observed radial velocity of S2 star is affected by both special relativistic and GR effects. The frequency shift measured by the observer is defined as
\begin{equation}
\zeta = \frac{\Delta\nu}{\nu} = \frac{\nu_{\rm em}-\nu_{\rm obs}}{\nu_{\rm obs}} = \frac{V_R}{c},
\end{equation}
where $\nu_{\rm em}$ and $\nu_{\rm obs}$ are the emission and observed photon frequencies, respectively, and $V_R$ is the line-of-sight velocity of the star. The total frequency shift includes two contributions: special relativistic Doppler shift, caused by the star's motion relative to the observer,
\begin{equation}
\zeta_D = \frac{\sqrt{1 - \mathsf{v_{\rm em}^2/c^2}}}{1 - \vec{n}\cdot \vec{\mathsf{v}}_{\rm em}/c},
\end{equation}
where $\mathsf{v_{\rm em}}$ is the orbital velocity at emission time, and $\vec{n}\cdot \vec{\mathsf{v}}_{\rm em}$ represents the velocity component along the line of sight.
And gravitational redshift, a GR effect arising from the gravitational potential at the point of emission,
\begin{equation}
\zeta_G = \frac{1}{\sqrt{|g_{tt}|}},
\end{equation}
evaluated along the star's worldline.
Combining these contributions yields the total relativistic frequency shift
\begin{equation}
\zeta = \zeta_D \cdot \zeta_G - 1 .
\end{equation}

Finally, the radial velocity as measured in the observer's frame must also be corrected for the possible motion of the central mass Sgr~A*. Any residual motion either offset or drift along the line of sight induces an additional velocity component. To model this, we introduce a parameter $v_{z0}$ in analogy with the sky-plane corrections. The full observed radial velocity becomes \cite{astro-ph/0612164}
\begin{equation}
\label{eq:total_radial_V}
V_R = c \cdot \zeta + \mathsf{v_{z0}}.
\end{equation}

In order to compare theoretical predictions with observational data, we analyze the orbital motion of the S2 star as predicted by the Schwarzschild-like black hole \eqref{eq:bumblebee_metric_sec2} and  Schwarzschild black hole. Fig.~\ref{fig:OBC}  presents the orbit of S2 star (left panel) and  the variation of radial velocity over time (right panel) for each model. The results show that the theoretical models are consistent with current observational data, and that the differences between these models are very small.

%%%%%%%%%%%%
%%%%%%%%%%%%
\section{Simulation with Monte Carlo Markov Chain} \label{sec:MCMC}
With the previous preparations, we are ready to perform an MCMC analysis to constrain the Lorentz-violating parameter $\ell$ in the model. For this purpose, we use the widely adopted Python package emcee \cite{Foreman-Mackey:2012any}, which implements an affine-invariant ensemble sampler particularly well suited for exploring high-dimensional and potentially correlated parameter spaces. By combining the astrometric positions, radial velocities, and the measured pericentre precession of S2 star within a unified likelihood framework, the MCMC procedure allows us to obtain statistically robust constraints on $\ell$ and to assess its compatibility with the current observational data.

\subsection{Analysis of MCMC}
The full set of parameters sampled in our MCMC analysis is
\begin{equation}
\label{eq:parameter}
\Big\{ M,R_{0},a,e,\iota,\omega,\Omega,t_{\text{apo}},\mathsf{x_0},\mathsf{y_{0}},\mathsf{v_{x0}},\mathsf{v_{y0}},\mathsf{v_{z0}}, \ell \Big\}.
\end{equation}
Here, the first two parameters $\{M,R_{0}\}$ denote the mass of the central SMBH Sgr~A*, and its distance from Earth, respectively. The next six parameters $\{a,e,\iota,\omega,\Omega,t_{\text{apo}}\}$ describe Keplerian elements of the S2 star's orbit: semi-major axis, eccentricity, orbital inclination, argument of pericentre, longitude of the ascending node, and the time of apoapsis passage. The following five parameters $\{\mathsf{x_0},\mathsf{y_{0}},\mathsf{v_{x0}},\mathsf{v_{y0}},\mathsf{v_{z0}}\}$ quantify the astrometric zero-point offsets and linear drifts of the reference frame with respect to the mass centroid. Finally, the last parameter $\{\ell\}$ is the Lorentz-violating parameter, whose value is the primary target of this work.

As we aforementioned, three different components of data are employed in our MCMC analysis. Accordingly, the total likelihood function is constructed as the product of three contributions,
\begin{equation}
\log \mathcal{L} =\log \mathcal{L}_{AP}+\log \mathcal{L}_{V_R}+\log \mathcal{L}_{\Delta\phi}.
\end{equation}
Here $\log \mathcal{L}_{AP}$ calculates the log-likelihood of the  145 astrometric positional data  of S2,
\begin{equation} \label{eq:AP}
\log \mathcal{L}_{AP} = -\frac{1}{2}\sum_{i} \frac{(X_{\text{obs}}^i-X_{\text{the}}^i)^2}{(\sigma^i_{X,\text{obs}})^2}-\frac{1}{2}\sum_{i} \frac{(Y_{\text{obs}}^i-Y_{\text{the}}^i)^2}{(\sigma^i_{Y,\text{obs}})^2} ,
\end{equation}
while $\log \mathcal{L}_{V_R}$ calculates the log-likelihood function values for the 44 radial velocity data,
\begin{equation} \label{eq:VR}
\log \mathcal{L}_{V_R} = -\frac{1}{2}\sum_{i} \frac{(V_{R,\text{obs}}^i-V^i_{R,\text{the}})^2}{(\sigma^i_{V_{R, \text{obs}}})^2} ,
\end{equation}
and $\log \mathcal{L}_{\Delta\phi}$ is the log-likelihood function values for the orbital precession,
\begin{equation} \label{eq:lP}
\log \mathcal{L}_{\Delta\phi} = -\frac{1}{2}\frac{(f_{\rm sp,\text{obs}}-f_{\rm sp,\text{the}})^2}{\sigma^2_{f_{\rm sp,\text{obs}}}} .
\end{equation}
In the above equations \eqref{eq:AP}-\eqref{eq:lP}, $X^i_{\text{obs}}$, $Y^i_{\text{obs}}$, $V^i_{R,\text{obs}}$ denote the observed astrometric position and velocity of the S2 star at the $i-$th epoch. Correspondingly, $X_{\text{the}}^i$, $Y_{\text{the}}^i$, $V^i_{R, \text{the}}$ are the theoretically predicted values computed from the orbital dynamics in above black holes. For the additional constraint introduced by the orbital precession, $f_{sp,\text{obs}}$ represents the observationally inferred value of the spin-precession (or apsidal-precession) factor present in \eqref{eq:deltaPHI} while $f_{sp, \text{the}} \equiv \Delta\phi/\Delta\phi_{GR}$ where $\Delta\phi$ is calculated in \eqref{eq:bum_orbit_period}. In each likelihood, $\sigma$ denotes the statistical uncertainty associated with the relevant measurement.

\subsection{Results}
We then carry out an MCMC simulation of the full 14-dimensional parameter space using the observational orbital data of the S2 star around the Galactic Center. For comparison, we consider two sets of priors adopted for the parameters, including the uniform prior and Gaussian prior, which are summarized in Table~\ref{tab:table01}. In both cases, we consider the uniform prior within $\ell \in [-0.9, 0.9]$ for the Lorentz-violating parameter.

\begin{table}[h]
\centering
\setlength{\tabcolsep}{10pt}
\renewcommand{\arraystretch}{1.2}
\begin{tabular}{|l|c|c|c|c|}
\hline
\multirow{2}{*}{Parameter} & \multicolumn{2}{c|}{Uniform prior} & \multicolumn{2}{c|}{Gaussian prior} \\
\cline{2-5}
 & Start & End & $\mu$ & $\sigma$ \\
\hline
$a$ (mas)        & 123       & 133    & 125.058   & 0.041     \\
$e$              & 0.5       & 0.9    & 0.884649  & 0.000066  \\
$M$ ($10^6 M_{\bigodot}$) & 2.0 & 6.5   & 4.261    & 0.0012   \\
$t_{\text{apo}}$ (yr)       & 1994.0    & 1994.8   & 1994.2913 & 0.0016   \\
$\mathsf{x_0}$ (mas)      & -1        & 1         & -0.9     & 0.14     \\
$\mathsf{y_0}$ (mas)      & -1        & 1         & 0.07     & 0.12     \\
$\iota$ ($^\circ$)   & 132.0     & 136.0    & 134.567  & 0.033    \\
$\omega$ ($^\circ$)  & 65.0      & 68.7     & 66.263   & 0.031    \\
$\Omega$ ($^\circ$)  & 217.0     & 230.5    & 228.171  & 0.031    \\
$R_0$ (kpc)          & 4         & 12       & 8.2467   & 0.0093   \\
$\mathsf{v_{x0}}$ (mas/yr)   & -1.0      & 1.0      & 0.080    & 0.010    \\
$\mathsf{v_{y0}}$ (mas/yr)   & -1.0      & 1.0      & 0.0341   & 0.00096  \\
$\mathsf{v_{z0}}$ (km/s)     & -50       & 50       & -1.6     & 1.4      \\
\hline
$\ell$              & \multicolumn{4}{c|}{Uniform prior within $[-0.9,0.9]$} \\
\hline
\end{tabular}
\caption{Two different sets of prior used for our MCMC analysis. The Gaussian prior data are taken from \cite{GRAVITY:2020gka}. In this table, $kpc$ is the kiloparsec, $mas$ is the milliarcsecond, $\circ$ is the degree, and $yr$ is the year.}
\label{tab:table01}
\end{table}

\begin{figure} [h]
{\centering
\includegraphics[width=6.5in]{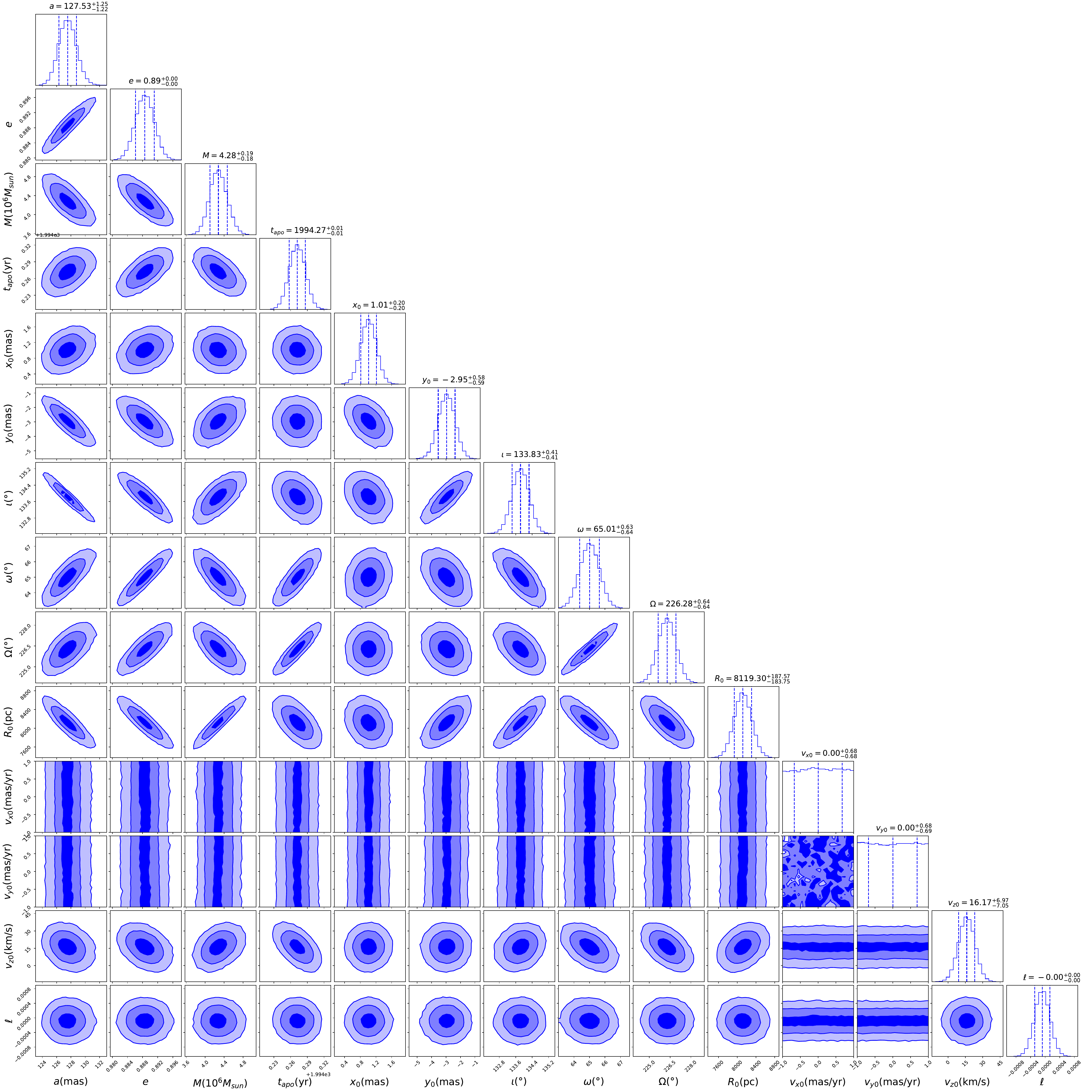} \hspace{0.5cm}
   \caption{ The posterior distributions of the 14 parameters under uniform priors listed in Table \ref{tab:table01}. The dashed lines mark values of the best fit and $1\sigma$ confidence levels, which are also collected in Table \ref{tab:table02}.}   \label{fig:Bum_uni}}
\end{figure}

\begin{figure} [h]
{\centering
\includegraphics[width=6.5in]{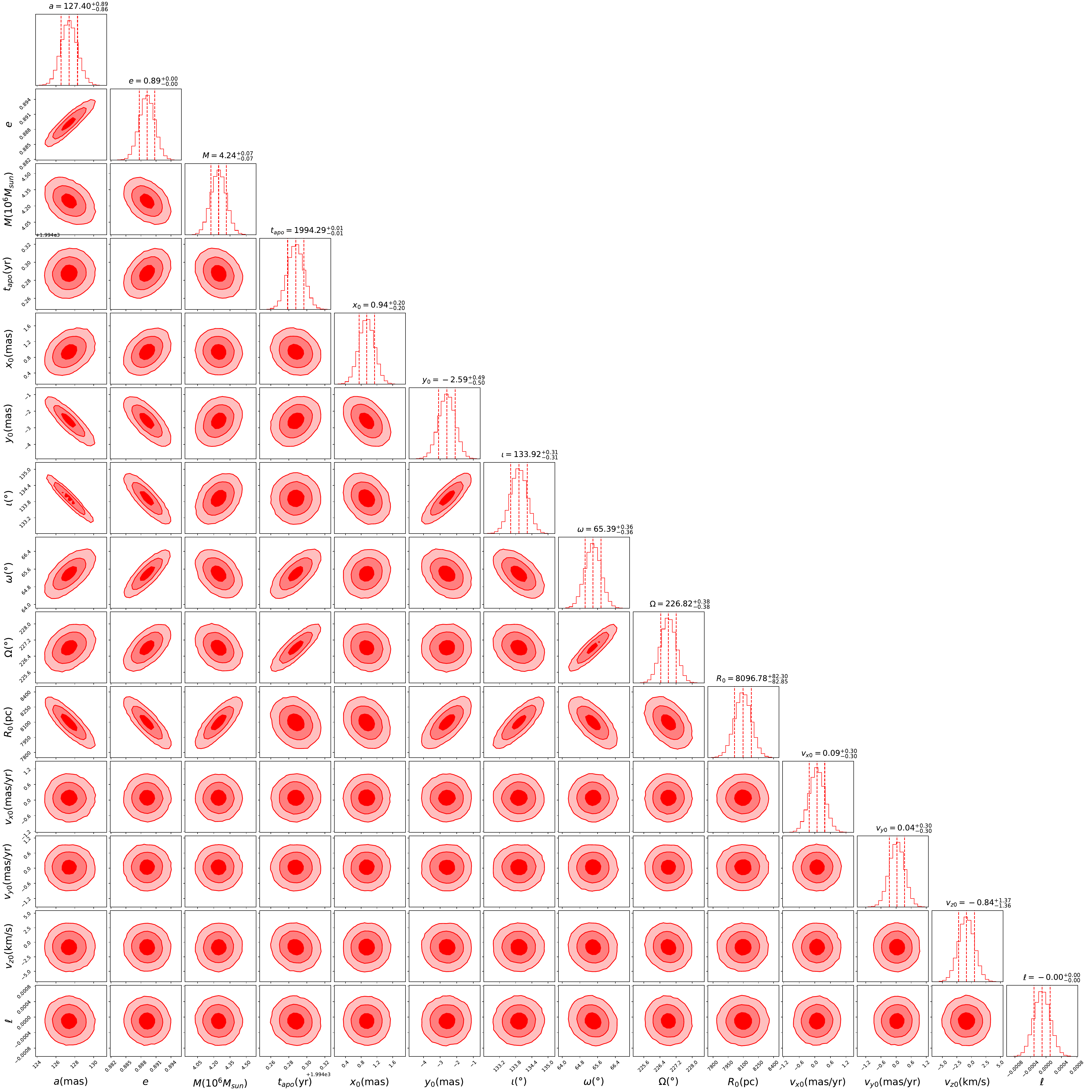} \hspace{0.5cm}
   \caption{ The posterior distributions of 14 parameters under Gaussian priors listed in Table \ref{tab:table01}. The dashed lines mark values of the best fit and $1\sigma$ confidence levels, which are also collected in Table \ref{tab:table02}.}   \label{fig:Bum_Gau}}
\end{figure}

The resulting posterior distributions of the orbital parameters, together with the Lorentz-violating  parameter $\ell$, are displayed in Figs.~\ref{fig:Bum_uni} and \ref{fig:Bum_Gau} under the uniform prior and Gaussian prior, respectively.
In each figure, the diagonal panels present the marginalized one-dimensional probability distributions, while the off-diagonal panels show the two-dimensional joint posterior contours for all parameter pairs.
The shaded regions in the contour plots correspond to the $1\sigma$, $2\sigma$ and $3\sigma$ confidence levels, respectively, thus illustrating the credible regions allowed by the current S2 observational dataset.

\begin{table}[h]
\centering
\setlength{\tabcolsep}{10pt}
\renewcommand{\arraystretch}{1.2}
\begin{tabular}{|l|c|c|}
\hline
Parameter & best-fit value (Uniform Prior) & best-fit value (Gaussian Prior) \\
\hline
$a$ (mas)          & $127.53^{+1.25}_{-1.22}$  & $127.40 ^{+0.89}_{-0.86}$  \\
$e$                 & $0.89 \pm 0$         & $0.89 \pm 0$        \\
$M$ ($10^6 M_{\text{Sun}}$)  & $4.28 ^{+0.19}_{-0.18}$    & $4.24 \pm 0.07$    \\
$t_{\text{apo}}$ (yr)       & $1994.27 \pm 0.01$      & $1994.29 \pm 0.01$   \\
$\mathsf{x_0}$ (mas)       & $1.01 \pm 0.20$        & $0.94 \pm 0.20$   \\
$\mathsf{y_0}$ (mas)     & $-2.95^{+0.58}_{-0.59}$   & $-2.59 ^{+0.49}_{-0.50}$  \\
$\iota$ ($^\circ$)     & $133.83 \pm 0.41$  & $133.92\pm 0.31$   \\
$\omega$ ($^\circ$)    & $65.01^{+0.63}_{-0.64}$   & $65.39 \pm 0.36$  \\
$\Omega$ ($^\circ$)    & $226.28 \pm 0.64$  & $226.82 \pm 0.38$ \\
$R_0$ (pc)     & $8119.30^{+187.57}_{-183.75}$ & $8096.78^{+82.30}_{-82.85}$ \\
$\mathsf{v_{x0}}$ (mas/yr)      & $0.00 \pm 0.68$   & $0.09 \pm 0.30 $    \\
$\mathsf{v_{y0}}$ (mas/yr)         & $0.00 ^{+0.68}_{-0.69}$   & $0.04 \pm 0.30 $  \\
$\mathsf{v_{z0}}$ (km/s)        & $16.17 ^{+6.97}_{-7.05}$   & $-0.84^{+1.37}_{-1.36}$   \\
$\ell$           & ${-8.01 \times 10^{-5}}^{+2.11 \times 10^{-4}}_{-2.09 \times 10^{-4}}$      & ${-1.00 \times 10^{-5}}^{+2.11 \times 10^{-4}}_{-2.09 \times 10^{-4}}$      \\
\hline
\end{tabular}
\caption{The best-fit values of the parameters of the orbital
model of S2 with two different priors. The uncertainties are given at the $1\sigma$ level.
\label{tab:table02} }
\end{table}

The corresponding best-fit values of all 14 parameters are summarized in Table~\ref{tab:table02}.
We now focus on the constraints on the Lorentz-violating parameter $\ell$ obtained under different prior assumptions. For the uniform prior, the Lorentz-violating parameter is constrained at the $1\sigma$ confidence level to be
\begin{equation}\label{eq:bound1}
\ell = {-8.01 \times 10^{-5}}^{+2.11 \times 10^{-4}}_{-2.09 \times 10^{-4}} .
\end{equation}
When the Gaussian prior is adopted, we obtain
\begin{equation}\label{eq:bound2}
\ell = {-1.00 \times 10^{-5}}^{+2.11 \times 10^{-4}}_{-2.09 \times 10^{-4}} .
\end{equation}
It is worthwhile pointing out that we also perform the MCMC simulation under the uniform prior within $\ell \in [-10^{-3},10^{-3}]$, and obtain the consistent constraints in both cases.

Since the metric \eqref{eq:bumblebee_metric_sec2} only has modifications to the
Schwarzschild spacetime appear in the $g_{rr}$ term, it does not change the size of the shadow comparing against Schwarzschild black hole. Therefore, the EHT observations of Sgr~A*, which connect
the size of the bright ring of emission to that of the underlying black hole shadow of Sgr~A*, cannot be used to constrain $\ell$ in the current model for SMBH \cite{Vagnozzi:2022moj}. However, for the rotating counterpart of the black hole  \eqref{eq:bumblebee_metric_sec2}, the EHT observations gave the constraint on $\ell$ is $\ell<\mathcal{O}(10^{-2})$ in  $1\sigma$ confidence region \cite{Kumar:2025bim}. Comparing against those constraints on the Lorentz violating parameter in SMBH scale,  the resultant constraint in our scenario is much stricter.

It should be noted that our constraints, \eqref{eq:bound1} and \eqref{eq:bound2}, are indeed weaker than those previously derived from the perihelion-shift measurements of Solar System planets \cite{Casana:2017jkc}. This is not unexpected because the orbital monitoring of S2 star is still limited by both instrumental precision and the complexity of modeling stellar dynamics in the dense environment near Sgr~A*. Nevertheless, despite its relatively larger uncertainties, the S2 dataset offers an independent and complementary probe of Lorentz-violating effects. In particular, the constraints obtained here are derived from a physical environment fundamentally different from the Solar System, namely, the strong-gravity regime in the immediate vicinity of a SMBH.
Thus, while the current limits from S2 star do not surpass Solar System bounds in absolute precision, they demonstrate that Galactic Center stellar dynamics already possess the sensitivity required to test modifications to GR and to place meaningful bounds on Lorentz-violating parameter in this scenario.
This highlights the important role of S-stars as probes of fundamental physics in a regime where relativistic effects are significantly stronger, and points to the possibility of substantially improved constraints as future high-precision astrometric and spectroscopic observations, such as  GRAVITY+ \cite{abuter2022first} and Extremely Large Telescope (ELT) \cite{Padovani:2023dxc}, become available.

%%%%%%%%%%%%
%%%%%%%%%%%%
\section{Closing remarks} \label{sec:conclusion}
In this work, we have utilized the astrometric and spectroscopic monitoring of the S2 star around Sgr~A* to test Lorentz symmetry breaking in the context of bumblebee gravity theory. By performing a comprehensive MCMC exploration of the full 14-dimensional parameter space under both uniform and Gaussian priors, we have obtained robust constraints on the Lorentz-violating parameter $\ell$. The results are shown in \eqref{eq:bound1} and \eqref{eq:bound2}
at $1\sigma$ confidence level, respectively, which show no significant deviation from GR within current observational uncertainties.

An important result of this analysis is that the S2 star based bounds on $\ell$ are significantly tighter than those inferred from current EHT imaging of Sgr~A* when applied to similar Lorentz-violating model in SMBH scale. The EHT constraint on the Lorentz-violating parameter, which exploit horizon-scale shadow morphology, is at the level of $\ell\lesssim\mathcal{O}(10^{-2})$ in the rotating scenarios \cite{Kumar:2025bim} while the static case cannot modify the black hole shadow, whereas the S2 star's orbital dynamics in the current study  constrain $\ell$ at the $\mathcal{O}(10^{-5})$ level. On the other hand, although Solar System tests (e.g., Shapiro's time-delay) provide the most stringent limits on Lorentz violation in the bumblebee gravity \cite{Casana:2017jkc}, the present work probes a completely different physical regime characterized by strong gravitational fields near a SMBH. This highlights the importance of Galactic Center stellar dynamics as an independent and complementary probe of fundamental physics.

Several promising directions can further enhance and broaden the scope of these tests.
One interesting direction is to extend the stellar populations and full dynamical modeling.
Beyond S2 star, the orbits of other short-period S-stars and potential discoveries of closer-in objects can provide additional independent constraints. Combining multi-star dynamics with self-consistent modeling of the extended mass distribution and relativistic perturbations will reduce systematic uncertainties and enhance the robustness of tests. In fact, using the datasets of S55 and S38 stars to constrain the model parameters have been carried out  in \cite{Shen:2023kkm,jovanovic2023constraints,Pietroni:2022cur,Borka:2022sot}.
While this work focuses on the Schwarzschild-like black hole in bumblebee gravity found in \cite{Casana:2017jkc}, so it is direct to carry out the parallel studies for other black holes, as we mentioned  in the introduction,  in the same theory, and check if they can give consistent constraint on the Lorentz violating parameter. Additionally,  there exists a rich landscape of Lorentz-violating theories within the SME and related effective field theory frameworks. Future analyses can also extend the current study to constrain other parameter sectors, thereby mapping out a more complete picture of Lorentz symmetry in gravitational sectors.
Another realistic direction to improve the usability of the S stars by GRAVITY is adding
in corrections due to the rotation of the SMBH. The Kerr-like black hole solution in the bumblebee model is still missing, though some efforts were made \cite{Ding:2019mal,Maluf:2022knd}.
Extending our analysis to the S-stars dynamic with a rotating SMBH, such as the slow-rotation approximation black hole \cite{Liu:2022dcn}, deserves further investigations.

\begin{acknowledgments}
We appreciate Dr. Qiang Wu for helpful corresponding. This work is partly supported by Natural Science Foundation of China under Grants Nos.12375054 and 12175192, the Postgraduate Research $\&$ Practice Innovation Program of 
Jiangsu Province under Grant No. $\text{KYCX25\_3924}$, and Yangzhou Science and Technology Planning Project in Jiangsu Province of China (YZ2025233).
\end{acknowledgments}

\bibliography{ref}
\bibliographystyle{utphys}

\end{document}